\documentclass[fleqn,usenatbib,useAMS]{mnras}
 \usepackage{graphicx}
 \usepackage{amsmath}
 \usepackage{amssymb}
 \usepackage[T1]{fontenc}
 \usepackage{ae,aecompl}
 \usepackage{txfonts}

 \title[Planet Nine and apsidal anti-alignment]
       {Finding Planet Nine: apsidal anti-alignment Monte Carlo results}

 \author[C. de la Fuente Marcos and R. de la Fuente Marcos]
        {C.~de~la~Fuente~Marcos\thanks{E-mail: carlosdlfmarcos@gmail.com}
         and
         R. de la Fuente Marcos \\
         Apartado de Correos 3413, E-28080 Madrid, Spain}
 \date{Accepted 2016 July 19.
       Received 2016 July 13;
       in original form 2016 March 29}
 \pubyear{2016}
 
\begin{document}
  \label{firstpage}
  \pagerange{\pageref{firstpage}--\pageref{lastpage}}
  \maketitle

  \begin{abstract}
     The distribution of the orbital elements of the known extreme 
     trans-Neptunian objects or ETNOs has been found to be statistically 
     incompatible with that of an unperturbed asteroid population 
     following heliocentric or, better, barycentric orbits. Such trends, if
     confirmed by future discoveries of ETNOs, strongly suggest that one
     or more massive perturbers could be located well beyond Pluto. Within
     the trans-Plutonian planets paradigm, the Planet Nine hypothesis has
     received much attention as a robust scenario to explain the observed 
     clustering in physical space of the perihelia of seven ETNOs which 
     also exhibit clustering in orbital pole position. Here, we revisit the 
     subject of clustering in perihelia and poles of the known ETNOs using 
     barycentric orbits, and study the visibility of the latest incarnation 
     of the orbit of Planet Nine applying Monte Carlo techniques and 
     focusing on the effects of the apsidal anti-alignment constraint. We 
     provide visibility maps indicating the most likely location of this 
     putative planet if it is near aphelion. We also show that the 
     available data suggest that at least two massive perturbers are 
     present beyond Pluto. 
  \end{abstract}


  \section{Introduction}
     The distribution of the orbital parameters of the known extreme trans-Neptunian objects or ETNOs is statistically incompatible with 
     that of an unperturbed asteroid population following Keplerian orbits (de la Fuente Marcos \& de la Fuente Marcos 2014, 2016b; Trujillo 
     \& Sheppard 2014; de la Fuente Marcos, de la Fuente Marcos \& Aarseth 2015, 2016; Gomes, Soares \& Brasser 2015; Batygin \& Brown 2016; 
     Brown \& Batygin 2016; Malhotra, Volk \& Wang 2016). A number of plausible explanations have been suggested. These include the possible 
     existence of one (Trujillo \& Sheppard 2014; Gomes et al. 2015; Batygin \& Brown 2016; Brown \& Batygin 2016; Malhotra et al. 2016) or 
     more (de la Fuente Marcos \& de la Fuente Marcos 2014, 2016b; de la Fuente Marcos et al. 2015, 2016) trans-Plutonian planets, capture 
     of ETNOs within the Sun's natal open cluster (J\'{\i}lkov\'a et al. 2015), stellar encounters (Brasser \& Schwamb 2015; Feng \& 
     Bailer-Jones 2015), being a by-product of Neptune's migration (Brown \& Firth 2016) or the result of the inclination instability 
     (Madigan \& McCourt 2016), and having been induced by Milgromian dynamics (Pau\v{c}o \& Kla\v{c}ka 2016). 

     At present, most if not all of the unexpected orbital patterns found for the known ETNOs seem to be compatible with the trans-Plutonian 
     planets paradigm that predicts the presence of one or more planetary bodies well beyond Pluto. Within this paradigm, the best studied 
     theoretical framework is that of the so-called Planet Nine hypothesis, originally suggested by Batygin \& Brown (2016) and further 
     developed in Brown \& Batygin (2016). The goal of this analytically and numerically supported conjecture is not only to explain the 
     observed clustering in physical space of the perihelia and the positions of the orbital poles of seven ETNOs (see Appendix A for 
     further discussion), but also to account for other, previously puzzling, pieces of observational evidence like the existence of low 
     perihelion objects moving in nearly perpendicular orbits. The Planet Nine hypothesis is compatible with existing data (Cowan, Holder \& 
     Kaib 2016; Fienga et al. 2016; Fortney et al. 2016; Ginzburg, Sari \& Loeb 2016; Linder \& Mordasini 2016) but, if Planet Nine exists, 
     it cannot be too massive or bright to have escaped detection during the last two decades of surveys and astrometric studies (Luhman 
     2014; Cowan et al. 2016; Fienga et al. 2016; Fortney et al. 2016; Ginzburg et al. 2016; Linder \& Mordasini 2016). A super-Earth in the 
     sub-Neptunian mass range is most likely and such planet may have been scattered out of the region of the Jovian planets early in the 
     history of the Solar system (Bromley \& Kenyon 2016) or even captured from another planetary system (Li \& Adams 2016; Mustill, Raymond 
     \& Davies 2016); super-Earths may also form at 125--750 au from the Sun (Kenyon \& Bromley 2015, 2016).  

     The analysis of the visibility of Planet Nine presented in de la Fuente Marcos \& de la Fuente Marcos (2016a) revealed probable 
     locations of this putative planet based on data provided in Batygin \& Brown (2016) and Fienga et al. (2016); the original data have 
     been significantly updated in Brown \& Batygin (2016). In addition, independent calculations (de la Fuente Marcos et al. 2016) show 
     that the apsidal anti-alignment constraint originally discussed in Batygin \& Brown (2016) plays a fundamental role on the dynamical 
     impact of a putative Planet Nine on the orbital evolution of the known ETNOs. Here, we improve the results presented in de la Fuente 
     Marcos \& de la Fuente Marcos (2016a) focusing on the effects of the apsidal anti-alignment constraint. This paper is organized as 
     follows. Section~2 presents an analysis of clustering in barycentric elements, pericentre and orbital pole positions, which is 
     subsequently discussed. An updated evaluation of the visibility of Planet Nine virtual orbits at aphelion is given in Section~3. 
     Conclusions are summarized in Section~4. 

%
%
      \begin{table}
        \centering
        \fontsize{8}{11pt}\selectfont
        \tabcolsep 0.16truecm
        \caption{Pericentre distances, $q$, ecliptic coordinates at pericentre, $(L_q, B_q)$, and projected pole positions, $(L_{\rm p}, 
                 B_{\rm p})$, of the 16 objects discussed in this paper computed using barycentric orbits, see also Figs \ref{cluster} and
                 \ref{poles}. (Epoch: 2457600.5, 2016 July 31.0 00:00:00.0 TDB. J2000.0 ecliptic and equinox. Input data from the SBDB; data 
                 as of 2016 July 13.)
                }
        \begin{tabular}{lccccc}
          \hline
             Object                   & $q$ (au) & $L_q$ (\degr) & $B_q$ (\degr) & $L_{\rm p}$ (\degr) & $B_{\rm p}$ (\degr) \\
          \hline
              (82158) 2001 FP$_{185}$ & 34.25    & 185.28        &     3.52      &  89.36              &    59.20            \\
              (90377) Sedna           & 76.19    &  96.31        &  $-$8.94      &  54.40              &    78.07            \\
             (148209) 2000 CR$_{105}$ & 44.12    &  87.28        & $-$15.39      &  38.29              &    67.24            \\
             (445473) 2010 VZ$_{98}$  & 34.35    &  71.21        &  $-$3.26      &  27.40              &    85.49            \\
             2002~GB$_{32}$           & 35.34    & 213.24        &     8.49      &  87.04              &    75.81            \\
             2003~HB$_{57}$           & 38.10    & 208.32        &     2.88      & 107.87              &    74.50            \\
             2003~SS$_{422}$          & 39.42    & 359.91        &  $-$8.28      &  61.05              &    73.21            \\
             2004~VN$_{112}$          & 47.32    &  35.65        & $-$13.59      & 336.02              &    64.45            \\
             2005~RH$_{52}$           & 39.00    & 336.98        &    10.83      & 216.11              &    69.55            \\
             2007~TG$_{422}$          & 35.56    &  39.41        & $-$17.88      &  22.91              &    71.40            \\
             2007~VJ$_{305}$          & 35.18    &   3.15        &  $-$4.40      & 294.38              &    78.02            \\
             2010~GB$_{174}$          & 48.56    & 118.83        &  $-$4.65      &  40.71              &    68.44            \\
             2012~VP$_{113}$          & 80.44    &  26.32        & $-$21.94      &   0.80              &    65.95            \\
             2013~GP$_{136}$          & 41.06    & 248.08        &    21.91      & 120.73              &    56.46            \\
             2013~RF$_{98}$           & 36.28    &  27.88        & $-$19.93      & 337.53              &    60.40            \\
             2015~SO$_{20}$           & 33.17    &  28.89        &  $-$2.05      & 303.63              &    66.59            \\
          \hline
        \end{tabular}
        \label{peri}
      \end{table}
%
%
%
%
      \begin{figure}
        \centering
         \includegraphics[width=\linewidth]{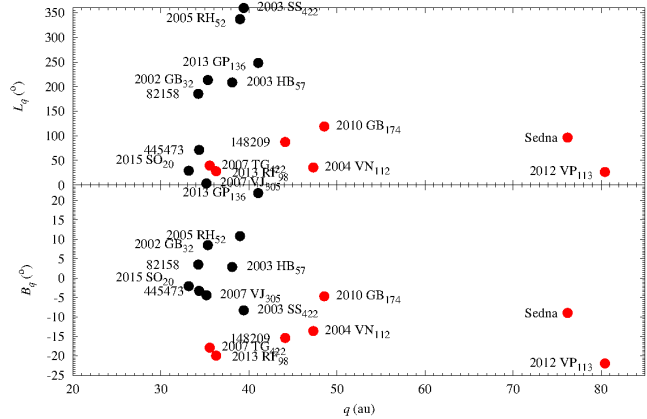}
         \caption{Pericentres of the objects in Table \ref{peri}. The objects singled out in Brown \& Batygin (2016) are plotted in red.  
                 }
         \label{cluster}
      \end{figure}
%
%
%
%
      \begin{figure}
        \centering
         \includegraphics[width=\linewidth]{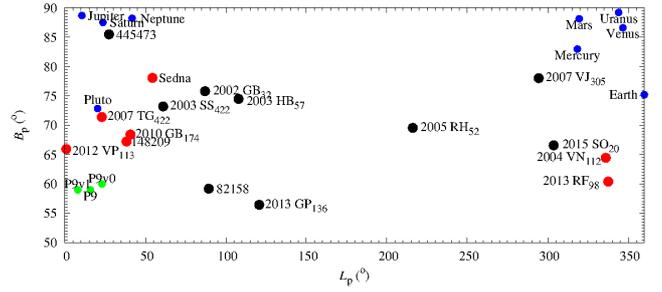}
         \caption{Poles of the objects in Table \ref{peri}. Those singled out in Brown \& Batygin (2016) are plotted in red, the known 
                  planets in blue, and various Planet Nine incarnations in green ---P9v0 is the nominal solution in Batygin \& Brown (2016), 
                  P9v1 is the one from Brown \& Batygin (2016), and P9 is the previous one but enforcing apsidal anti-alignment.  
                 }
         \label{poles}
      \end{figure}
%
%
%
%
      \begin{table*}
        \centering
        \fontsize{8}{11pt}\selectfont
        \tabcolsep 0.10truecm
        \caption{Barycentric orbital elements and parameters ---$q=a(1-e)$, $Q=a(1+e)$ is the aphelion distance, $\varpi=\Omega+\omega$, $P$ 
                 is the orbital period, $\Omega^*$ and $\omega^*$ are $\Omega$ and $\omega$ in the interval ($-\pi$, $\pi$) instead of the 
                 regular (0, 2$\pi$)--- of the known ETNOs. The statistical parameters are Q$_{1}$, first quartile, Q$_{3}$, third quartile, 
                 IQR, interquartile range, OL, lower outlier limit (Q$_{1}-1.5$IQR), and OU, upper outlier limit (Q$_{3}+1.5$IQR). Input 
                 data as in Table \ref{peri}.
                }
        \begin{tabular}{lrrrrrrrrrrr}
          \hline
             Object             & $a$ (au)  & $e$     & $i$ (\degr) & $\Omega$ (\degr) & $\omega$ (\degr) & $\varpi$ (\degr) & $q$ (au) & 
                       $Q$ (au) & $P$ (yr)    & $\Omega^*$ (\degr) & $\omega^*$ (\degr) \\
          \hline
                          82158 & 215.97915 & 0.84141 & 30.80134    & 179.35892        &   6.88451        & 186.24343        & 34.25244 &
                      397.70586 &  3172.01164 &    179.35892       &     6.88451        \\
                          Sedna & 506.08846 & 0.84945 & 11.92856    & 144.40251        & 311.28569        &  95.68820        & 76.19098 &
                      935.98594 & 11377.75735 &    144.40251       &  $-$48.71431       \\
                         148209 & 221.97188 & 0.80122 & 22.75598    & 128.28590        & 316.68922        &  84.97512        & 44.12267 &
                      399.82108 &  3304.94285 &    128.28590       &  $-$43.31078       \\
                         445473 & 153.36100 & 0.77602 &  4.51050    & 117.39858        & 313.72557        &  71.12415        & 34.35048 &
                      272.37152 &  1897.97030 &    117.39858       &  $-$46.27443       \\
                2002 GB$_{32}$  & 206.50931 & 0.82887 & 14.19246    & 177.04395        &  37.04720        & 214.09115        & 35.33979 &
                      377.67883 &  2965.69498 &    177.04395       &     37.04720       \\
                2003 HB$_{57}$  & 159.66557 & 0.76138 & 15.50028    & 197.87107        &  10.82977        & 208.70084        & 38.09895 &
                      281.23218 &  2016.20147 & $-$162.12893      &     10.82977       \\
                2003 SS$_{422}$ & 197.89567 & 0.80078 & 16.78597    & 151.04690        & 209.92864        &   0.97554        & 39.42455 &
                      356.36679 &  2782.09180 &    151.04690      & $-$150.07136       \\
                2004 VN$_{112}$ & 327.43521 & 0.85548 & 25.54761    &  66.02280        & 326.99699        &  33.01979        & 47.32201 &
                      607.54841 &  5921.13675 &     66.02280      &  $-$33.00301       \\
                2005 RH$_{52}$  & 153.67748 & 0.74624 & 20.44577    & 306.11067        &  32.53853        & 338.64920        & 38.99710 &
                      268.35786 &  1903.84838 &  $-$53.88933      &     32.53853       \\
                2007 TG$_{422}$ & 502.04248 & 0.92916 & 18.59530    & 112.91071        & 285.68512        &  38.59583        & 35.56265 &
                      968.52230 & 11241.58933 &    112.91071      &  $-$74.31488      \\
                2007 VJ$_{305}$ & 192.09934 & 0.81684 & 11.98376    &  24.38239        & 338.33491        &   2.71730        & 35.18470 &
                      349.01398 &  2660.76077 &     24.38239      &  $-$21.66509      \\
                2010 GB$_{174}$ & 351.12735 & 0.86169 & 21.56245    & 130.71444        & 347.24510        & 117.95954        & 48.56288 &
                      653.69182 &  6575.27641 &    130.71444      &  $-$12.75490      \\
                2012 VP$_{113}$ & 263.16564 & 0.69436 & 24.05155    &  90.80392        & 293.54965        &  24.35357        & 80.43515 &
                      445.89613 &  4266.39240 &     90.80392      &  $-$66.45035      \\
                2013 GP$_{136}$ & 149.78673 & 0.72587 & 33.53904    & 210.72729        &  42.47818        & 253.20547        & 41.06079 &
                      258.51267 &  1832.00656 & $-$149.27271      &     42.47818      \\
                2013 RF$_{98}$  & 317.06525 & 0.88557 & 29.60066    &  67.53381        & 316.37528        &  23.90909        & 36.28242 &
                      597.84808 &  5642.08982 &     67.53381      &  $-$43.62472      \\
                2015 SO$_{20}$  & 164.90289 & 0.79885 & 23.41106    &  33.63383        & 354.83023        &  28.46406        & 33.17008 &
                      296.63570 &  2116.21317 &     33.63383      &   $-$5.16977      \\
          \hline
             Mean               & 255.17334 & 0.81082 & 20.32577    & 133.64048        & 221.52654        & 107.66702        & 43.64735 &
                      466.69932 &  4354.74900 &     66.14048      &  $-$25.97346      \\ 
             Std. dev.          & 116.48633 & 0.06105 &  7.72495    &  71.95062        & 140.15770        & 102.40843        & 14.31074 &
                      226.78010 &  3103.33812 &    105.71229      &     49.06288      \\
             Median             & 211.24423 & 0.80903 & 21.00411    & 129.50017        & 302.41767        &  78.04963        & 38.54802 &
                      387.69235 &  3068.85331 &    101.85731      &  $-$27.33405      \\
             Q$_{1}$            & 163.59356 & 0.77236 & 15.17332    &  84.98639        &  41.12044        &  27.43644        & 35.30102 &
                      292.78482 &  2091.21024 &     31.32097      &  $-$46.88440      \\
             Q$_{3}$            & 319.65774 & 0.85096 & 24.42556    & 177.62269        & 319.26616        & 191.85778        & 44.92251 &
                      600.27317 &  5711.85155 &    134.13646      &      7.87082      \\
             IQR                & 156.06418 & 0.07860 &  9.25224    &  92.63630        & 278.14573        & 164.42135        &  9.62149 &
                      307.48835 &  3620.64131 &    102.81549      &     54.75523      \\
             OL                 & $-$70.50272 & 0.65446 &  1.29496 & $-$53.96806       &$-$376.09815      &$-$219.19558      & 20.86879 &
                   $-$168.44770 & $-$3339.75172 & $-$122.90226    & $-$129.01724      \\
             OU                 & 553.75402 & 0.96886 & 38.30393 & 316.57714           & 736.48475        & 438.48980        & 59.35474 &
                     1061.50568 & 11142.81351 &  288.35969        &   90.00366        \\
          \hline
             \multicolumn{12}{c}{Statistics of Sedna, 148209, 2004~VN$_{112}$, 2007~TG$_{422}$, 2010~GB$_{174}$, 2012~VP$_{113}$ and 2013~RF$_{98}$} \\
          \hline
             Mean               & 355.55661 & 0.83956 & 22.00602    & 105.81058        & 313.97529        &  59.78588        & 52.63982 &
                      658.47340 &  6904.16927 &    105.81058      &  $-$46.02471      \\ 
             Std. dev.          & 110.14455 & 0.07477 &  5.60287    &  31.46021        &  20.47089        &  38.76357        & 18.27509 &
                      220.42583 &  3199.11104 &     31.46021      &     20.47089      \\
             Median             & 327.43521 & 0.85548 & 22.75598    & 112.91071        & 316.37528        &  38.59583        & 47.32201 &
                      607.54841 &  5921.13675 &    112.91071      &  $-$43.62472      \\
             Q$_{1}$            & 290.11545 & 0.82534 & 20.07888    &  79.16886        & 302.41767        &  28.68668        & 40.20255 &
                      521.87211 &  4954.24111 &     79.16886      &  $-$57.58233      \\
             Q$_{3}$            & 426.58491 & 0.87363 & 24.79958    & 129.50017        & 321.84310        &  90.33166        & 62.37693 &
                      794.83888 &  8908.43287 &    129.50017      &  $-$38.15690      \\
             IQR                & 136.46947 & 0.04829 &  4.72070    &  50.33131        &  19.42543        &  61.64498        & 22.17438 &
                      272.96677 &  3954.19176 &     50.33131      &     19.42543      \\
             OL                 &  85.41124 & 0.75290 & 12.99782    &    3.67190       & 273.27952        & $-$63.78080      &  6.94097 &
                      112.42195 & $-$977.04653 &     3.67190      &  $-$86.72048      \\
             OU                 & 631.28912 & 0.94607 & 31.88063 & 204.99713           & 350.98125        & 182.79914        & 95.63851 &
                     1204.28903 & 14839.72051 &  204.99713        &   $-$9.01875      \\
          \hline
        \end{tabular}
        \label{bary}
      \end{table*}
%
%

  \section{Clustering in barycentric parameters}
     The six (Batygin \& Brown 2016) or seven (Brown \& Batygin 2016) ETNOs singled out within the Planet Nine hypothesis (see Appendix A 
     for details) have $a>226$~au (heliocentric) and they exhibit clustering in perihelion location in absolute terms and also in orbital 
     pole position. In order to better understand the context of these clusterings we study the line of apsides of the known ETNOs (see 
     Table \ref{peri} and Fig. \ref{cluster}) and the projection of their orbital poles on to the plane of the sky (see Table \ref{peri} and 
     Fig. \ref{poles}). Here, we consider barycentric orbits as it can be argued (see the discussion in Malhotra et al. 2016) that the use 
     of barycentric orbital elements instead of the usual heliocentric ones is more correct in this case. 

     In Trujillo \& Sheppard (2014), the ETNOs are defined as asteroids with heliocentric semimajor axis greater than 150 au and perihelion 
     greater than 30 au; at present, there are 16 known ETNOs. Because of the nature of their orbits, the ETNOs cannot experience a close 
     approach to the known planets, including Neptune. Nevertheless, the orbits of the ETNOs are affected by indirect perturbations that 
     induce variations in perihelion location. The perihelion distance of an ETNO depends on the value of its semimajor axis and 
     eccentricity, $e$, as it is given by $q=a(1-e)$. However, its absolute position on the sky is only function of the inclination, $i$, 
     the longitude of the ascending node, $\Omega$, and the argument of perihelion, $\omega$. In heliocentric ecliptic coordinates, the 
     longitude and latitude of an object at perihelion, $(l_q, b_q)$, are given by the expressions: $\tan{(l_q-\Omega)}=\tan\omega\,\cos{i}$ 
     and $\sin{b_q}=\sin\omega\,\sin{i}$ (see e.g. Murray \& Dermott 1999). For an orbit with zero inclination, $l_q=\Omega+\omega$ and 
     $b_q=0\degr$; therefore, if $i=0\degr$, the value of $l_q$ coincides with that of the longitude of perihelion parameter, 
     $\varpi=\Omega+\omega$. In Brown \& Batygin (2016), $l_q$ is called `perihelion longitude' and $b_q$ is the `perihelion latitude'. 
     Considering barycentric orbits, we denote the barycentric ecliptic coordinates of an ETNO at pericentre as $(L_q, B_q)$. In Table 
     \ref{peri} we show the values of $q$, $L_q$ and $B_q$ computed using the barycentric data in Table \ref{bary}. The input values are 
     from Jet Propulsion Laboratory's Small-Body Database\footnote{http://ssd.jpl.nasa.gov/sbdb.cgi} (SBDB) and \textsc{horizons} On-Line 
     Ephemeris System (Giorgini et al. 1996).

     Fig. \ref{cluster} shows the position in the sky at pericentre as a function of the pericentre distance for the known ETNOs. The 
     objects in Brown \& Batygin (2016) are plotted in red; the average angular separation at pericentre for this group is 
     44\degr$\pm$30\degr, but the pair 2012~VP$_{113}$--2013~RF$_{98}$ has a separation of 2\fdg5, 2004~VN$_{112}$--2007~TG$_{422}$ has 
     5\fdg6, and 2004~VN$_{112}$--2013~RF$_{98}$ has 9\fdg8. In addition to this clustering, other obvious groupings are also visible. The 
     dispersion of these additional groupings in ecliptic coordinates is similar to that of the set singled out by Brown \& Batygin (2016), 
     but their dispersion in $q$ is considerably lower. ETNOs (82158) 2001 FP$_{185}$, 2002~GB$_{32}$ and 2003 HB$_{57}$ have a nearly 
     common line of apsides; the same can be said about (445473) 2010 VZ$_{98}$, 2007~VJ$_{305}$ and 2015~SO$_{20}$. The case for the first 
     grouping is particularly strong. Out of 16 known ETNOs, only five reach pericentre north from the ecliptic, i.e. $B_q>0\degr$; these 
     objects are 82158, 2002~GB$_{32}$, 2003 HB$_{57}$, 2005 RH$_{52}$ and 2013~GP$_{136}$. This represents a 2$\sigma$ departure from an 
     isotropic distribution in $B_q$, where $\sigma=\sqrt{n}/2$ is the standard deviation for binomial statistics. This marginally 
     significant result suggests that an unknown massive perturber has aligned the apsidal orientation of these objects, but the Planet Nine 
     hypothesis cannot explain this pattern (Batygin \& Brown 2016); another perturber is needed as the line of apsides is scarcely affected 
     by indirect perturbations from the known planets. A nearly common line of apsides is also expected in a set of objects resulting from 
     the break-up of a single object at pericentre. Such scenario might be linked to some of the groupings observed. 

     In Table \ref{peri} we show the current values of the position in the sky of the poles of the orbits of the known ETNOs computed using 
     the data in Table \ref{bary}; the ecliptic coordinates of the pole are $(L_{\rm p}, B_{\rm p}) = (\Omega-90\degr, 90\degr-i)$. Fig. 
     \ref{poles} shows the poles of the known ETNOs and also those of the known planets of the Solar system and various nominal orbits of 
     Planet Nine (epoch 2457600.5 JD). The objects singled out in Brown \& Batygin (2016) and Planet Nine appear to exhibit a relative 
     arrangement in terms of position of their orbital poles similar to the one existing between Neptune and Pluto. For this group, the 
     average polar separation is 16\degr$\pm$8\degr, but the pair 148209--2010~GB$_{174}$ has 1\fdg5, 2004~VN$_{112}$--2013~RF$_{98}$ has 
     4\fdg1, 148209--2007~TG$_{422}$ has 6\fdg8, 2007~TG$_{422}$--2010~GB$_{174}$ has 6\fdg8, and 2007~TG$_{422}$--2012~VP$_{113}$ has 
     9\fdg6. On the other hand, the ETNOs 2002~GB$_{32}$ and 2003~HB$_{57}$ not only exhibit apsidal alignment but their orbital poles are 
     also very close. 

     As for the overall clustering in orbital parameter space, it has been claimed that the ETNOs exhibit clustering in the values of their 
     $\omega$ (Trujillo \& Sheppard 2014), $e$ and $i$ (de la Fuente Marcos \& de la Fuente Marcos 2014), and $\Omega$ (Brown \& Firth 
     2016). Table \ref{bary} presents the descriptive statistics of the known ETNOs; in this table, unphysical values are shown for 
     completeness. The bottom block of statistics in Table \ref{bary} (see also Appendix A) focuses on the seven ETNOs singled out in Brown 
     \& Batygin (2016). The overall statistics is only slightly different from that of the heliocentric orbital elements. The mean value of 
     the barycentric $e$ of the known ETNOs amounts to 0.81$\pm$0.06, the barycentric $i$ is 20\degr$\pm$8\degr, the barycentric $\Omega$ is 
     134\degr$\pm$72\degr, and the barycentric $\omega$ is $-$26\degr$\pm$49\degr (see $\omega^*$ in Table \ref{bary}). Clustering in $e$ 
     may be due to selection effects, but the others cannot be explained as resulting from observational biases, they must have a dynamical 
     origin (de la Fuente Marcos \& de la Fuente Marcos 2014). For the ETNO sample in Brown \& Batygin (2016), the average values of the 
     barycentric orbital parameters are $e=0.84\pm0.07$, $i=22\degr\pm6\degr$, $\Omega=106\degr\pm31\degr$ and $\omega=314\degr\pm20\degr$.

     Regarding the issue of statistical outliers and assuming that outliers are observations that fall below Q$_{1}-1.5$ IQR or above 
     Q$_{3}+1.5$ IQR, where Q$_{1}$ is the first quartile, Q$_{3}$ is the third quartile, and IQR is the interquartile range, we observe a 
     small but relevant number of outliers. For the entire sample, 2003 SS$_{422}$ is an outlier in terms of $\omega^*$ (see Table 
     \ref{bary}), Sedna and 2012 VP$_{113}$ are outliers in terms of $q$, and Sedna and 2007~TG$_{422}$ are outliers in terms of orbital 
     period. In terms of size (not shown in the table), Sedna is a very significant outlier with $H=1.6$~mag when the lower and upper limits 
     for outliers are 4.7~mag and 8.7~mag, respectively. As for the sample of ETNOs in Brown \& Batygin (2016), Sedna is an statistical 
     outlier in terms of $i$. 

  \section{Visibility analysis}
     Here, we apply the Monte Carlo approach (Metropolis \& Ulam 1949) described in de la Fuente Marcos \& de la Fuente Marcos (2014) to 
     construct visibility maps indicating the most probable location of this putative planet if it is near aphelion. Each Monte Carlo 
     experiment consists of $10^{7}$ test orbits uniformly distributed in parameter space. The analyses in e.g. Cowan et al. (2016), Fienga 
     et al. (2016), Fortney et al. (2016), Ginzburg et al. (2016) and Linder \& Mordasini (2016) strongly disfavour a present-day Planet 
     Nine located at perihelion and they do not discard the aphelion which is also favoured in Brown \& Batygin (2016). 

     \subsection{Batygin \& Brown (2016)}
        The resonant coupling mechanism described in Batygin \& Brown (2016) emphasizes the existence of simultaneous apsidal anti-alignment
        and nodal alignment, i.e. $\Delta\varpi$ librates about 180\degr and $\Delta\Omega$ librates about 0\degr. The relative values of 
        $\varpi$ and $\Omega$ of the ETNO with respect to those of the perturber must oscillate in order to maintain orbital confinement but
        see Beust (2016) for a detailed analysis. In Batygin \& Brown (2016), the value of $\omega$ of the putative Planet Nine is 
        138\degr$\pm$21\degr. It is also indicated that the average value of $\Omega$ for their six ETNOs is 113\degr$\pm$13\degr and that 
        of $\omega$ is 318\degr$\pm$8\degr. Applying the conditions for stability and using the other values discussed in their work, we 
        generate a synthetic population of Planet Nines with $a\in$ (400, 1500) au, $e\in$ (0.5, 0.8), $i\in$ (15, 45)\degr, $\Omega\in$ 
        (100, 126)\degr and $\omega\in$ (117, 159)\degr. Fig. \ref{hunt}, left-hand panels, shows the distribution in equatorial coordinates 
        of the set of studied orbits. In this figure, the value of the parameter in the appropriate units is colour coded following the 
        scale printed on the associated colour box. The location of the Galactic disc appears in panel D (inclination). The background 
        stellar density is the highest towards this region. The distribution of $Q$, $a$ and $e$ is rather uniform as expected because the 
        location in the sky of Planet Nine does not depend on these parameters (see above). The distribution in declination depends on $i$ 
        and $\omega$; those orbits with higher values of $i$ reach aphelion at lower declinations, the same behaviour is observed for the 
        ones with lower values of $\omega$. The location in right ascension mainly depends on the value of $\Omega$, both increasing in 
        direct proportion. Fig. \ref{radeca}, left-hand panels, shows that the frequency distribution in right ascension and declination is 
        far from uniform; the probability of finding an orbit reaching aphelion is highest in the region limited by $\alpha\in(4.5, 
        5.5)^{\rm h}$ and $\delta\in(6, 9)\degr$. However, the mean values of $\Omega$ and $\omega$ for the six ETNOs of interest differ 
        from those in Table \ref{bary2}.

     \subsection{Brown \& Batygin (2016)}
\begin{figure*}
        \centering
         \includegraphics[width=0.254\linewidth]{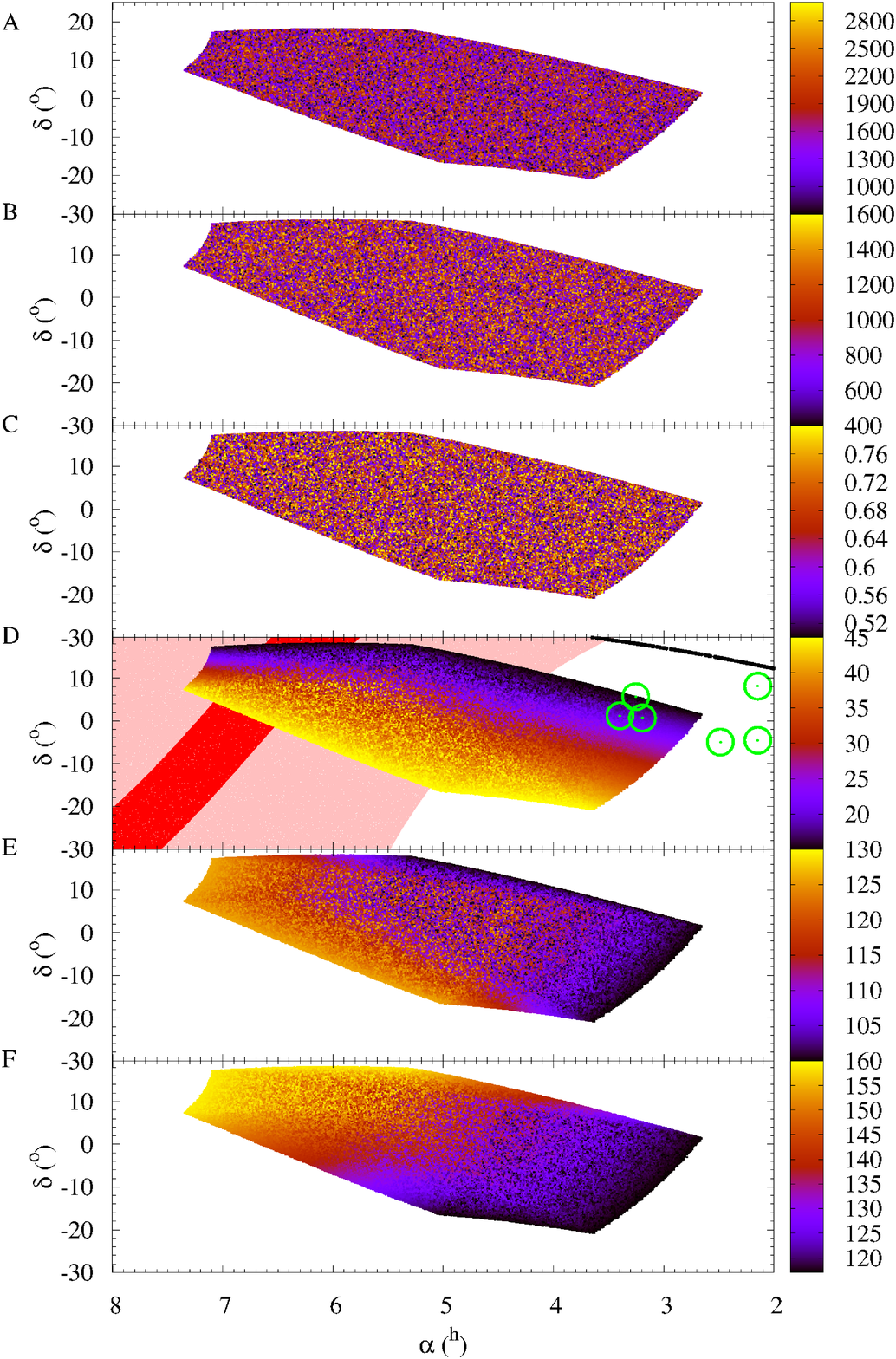}
         \includegraphics[width=0.244\linewidth]{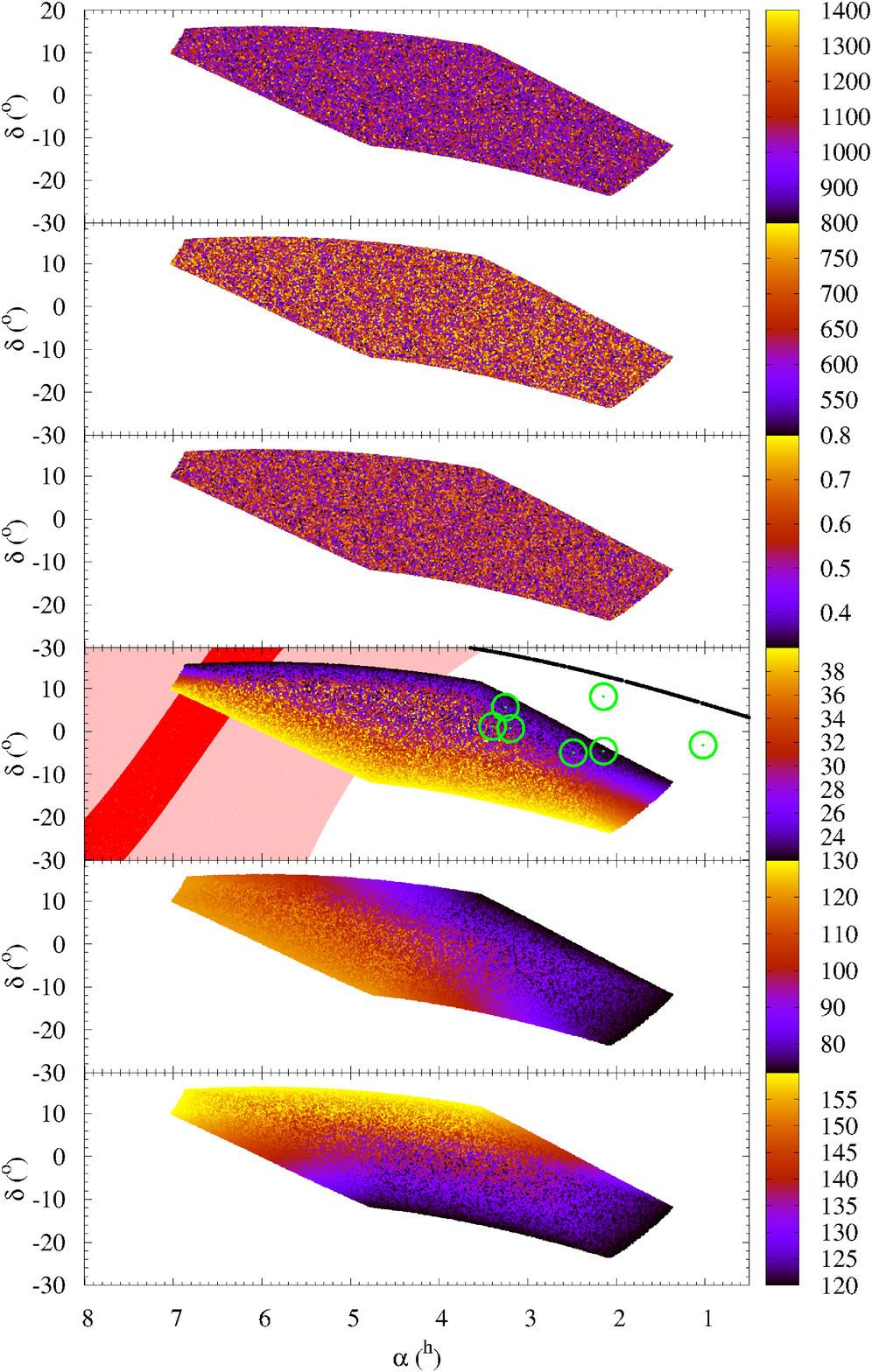}
         \includegraphics[width=0.244\linewidth]{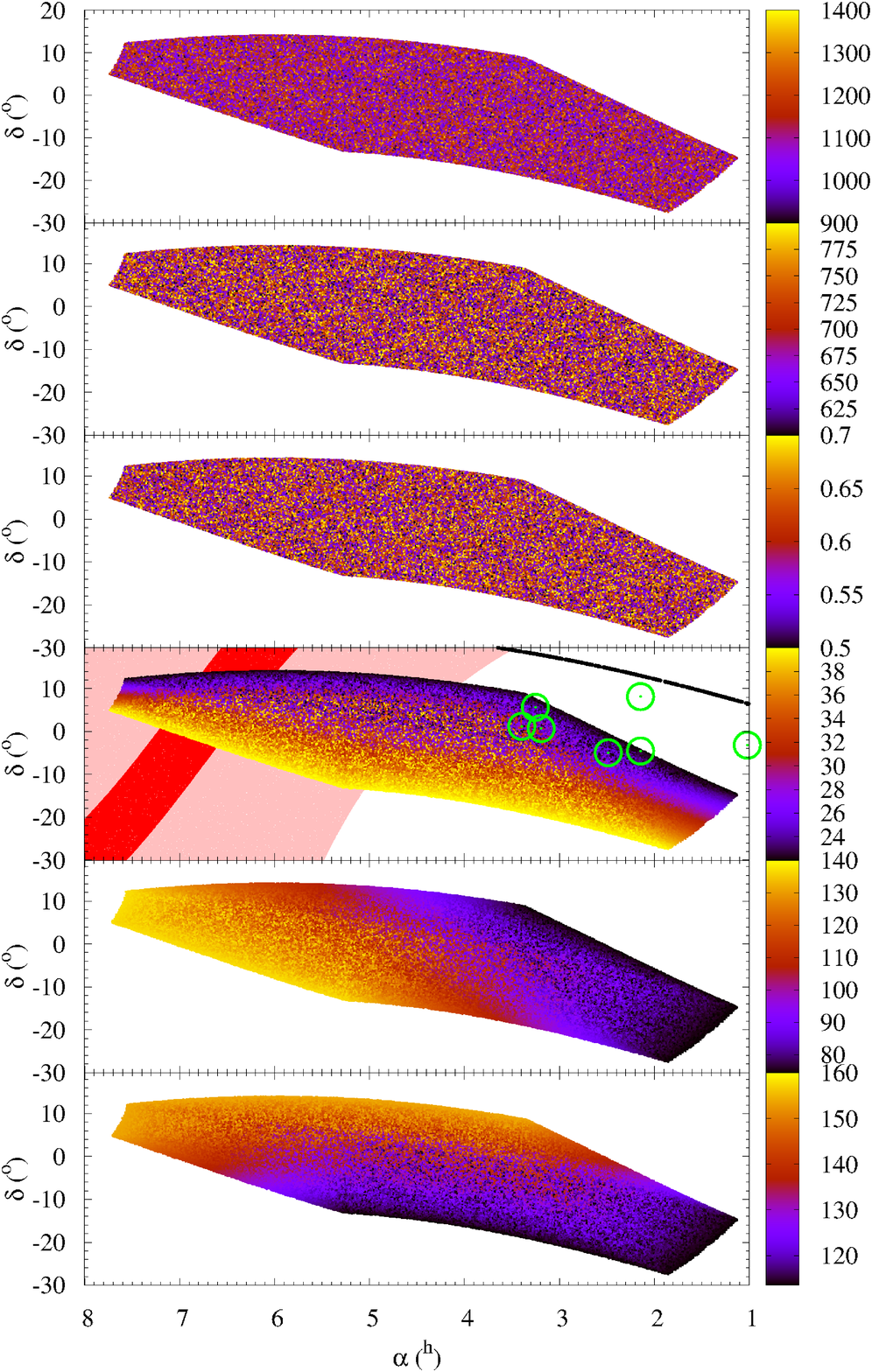}
         \includegraphics[width=0.244\linewidth]{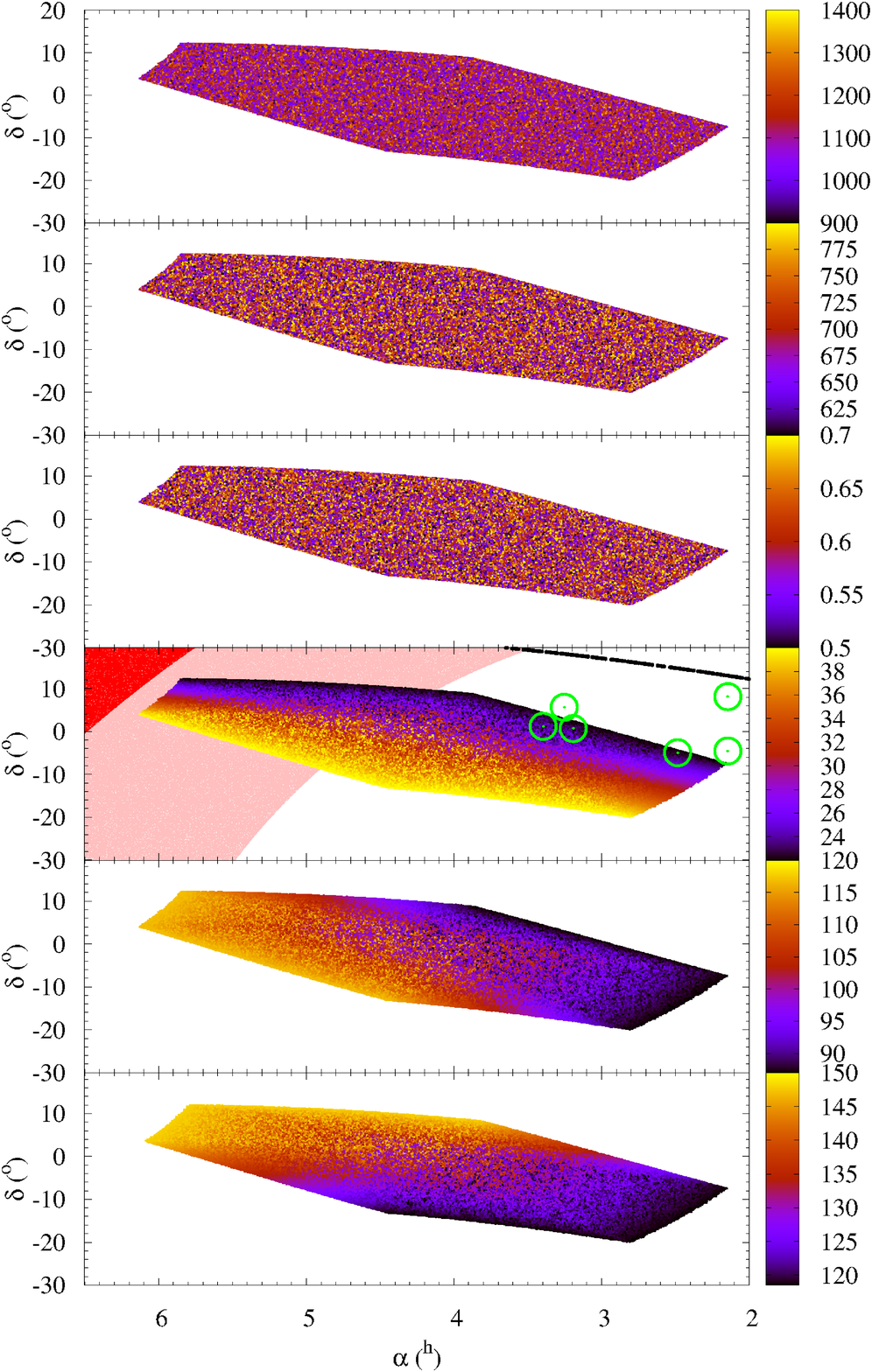}\\
         \caption{Distribution in equatorial coordinates of the aphelia of the studied orbits as a function of various orbital parameters: 
                  $Q$ (panel A), $a$ (panel B), $e$ (panel C), $i$ (panel D), $\Omega$ (panel E), and $\omega$ (panel F). The left-hand 
                  panels show results using the sets of orbits in Batygin \& Brown (2016), those of orbits from Brown \& Batygin (2016) are
                  displayed in the second to left-hand panels; the second to right-hand panels and the right-hand panels focus on the set 
                  of orbits described in Sect. 3.3, imposing $\Delta\varpi\sim$180\degr and $\Delta\Omega\sim$0\degr using the data in Table
                  \ref{bary} and \ref{bary2}, respectively. In panel D, the green circles give the location at discovery of known ETNOs (see
                  table 2 in de la Fuente Marcos \& de la Fuente Marcos 2016a), in red we have the Galactic disc that is arbitrarily defined 
                  as the region confined between galactic latitude $-5$\degr and 5\degr, in pink the region enclosed between galactic 
                  latitude $-30$\degr and 30\degr, and in black the ecliptic. 
                 }
         \label{hunt}
      \end{figure*}

\begin{figure*}
        \centering
         \includegraphics[width=0.247\linewidth]{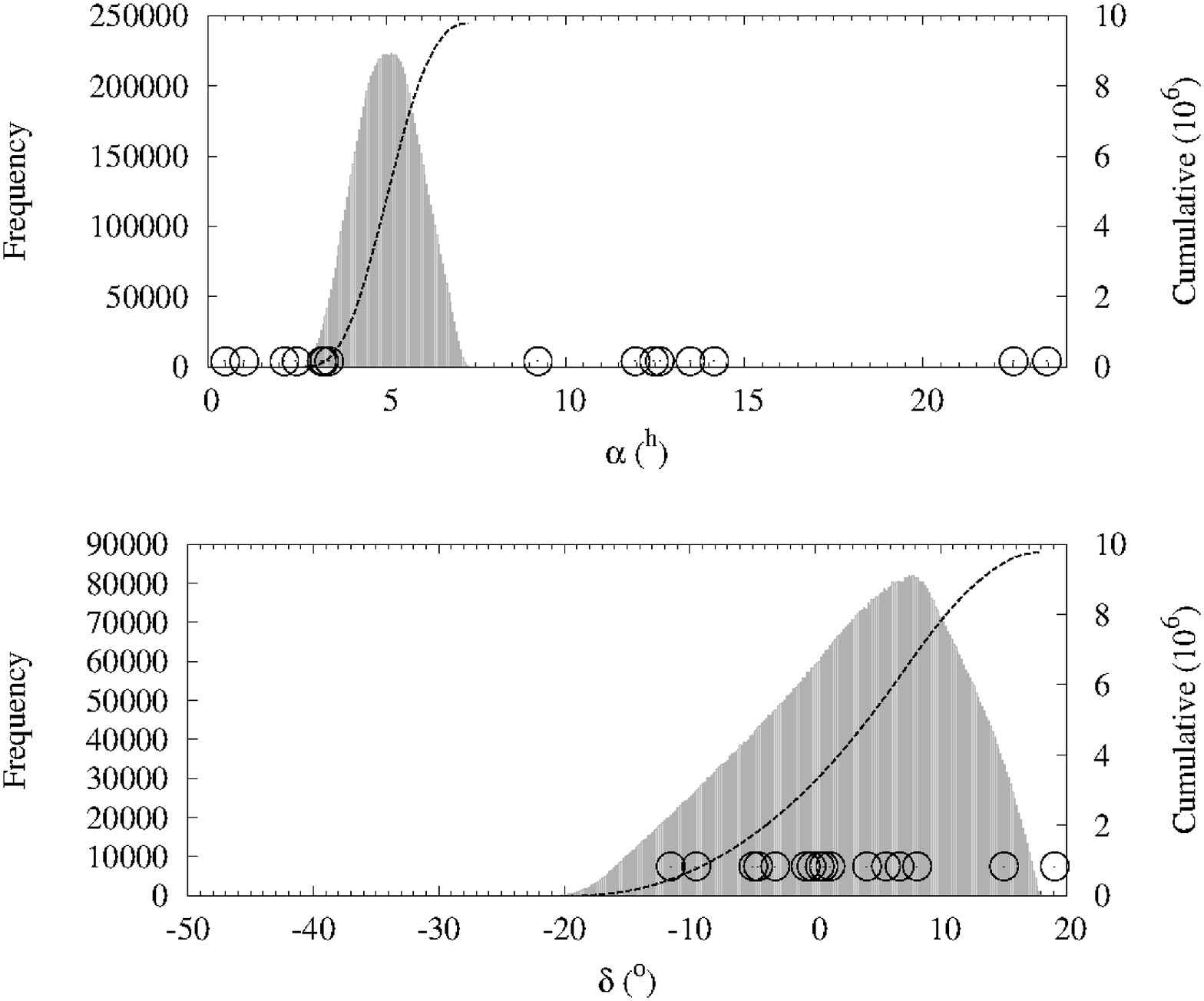}
         \includegraphics[width=0.247\linewidth]{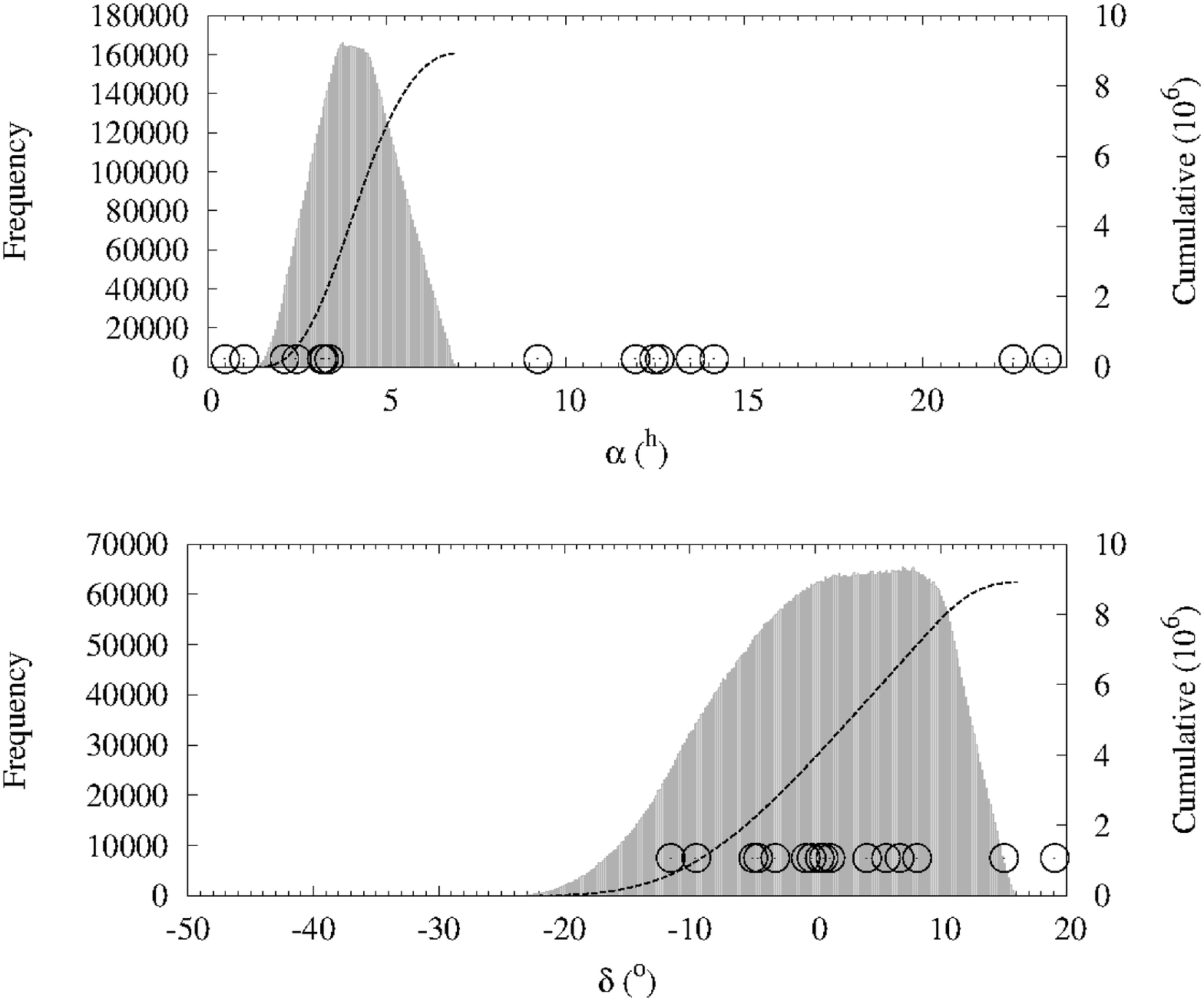}
         \includegraphics[width=0.247\linewidth]{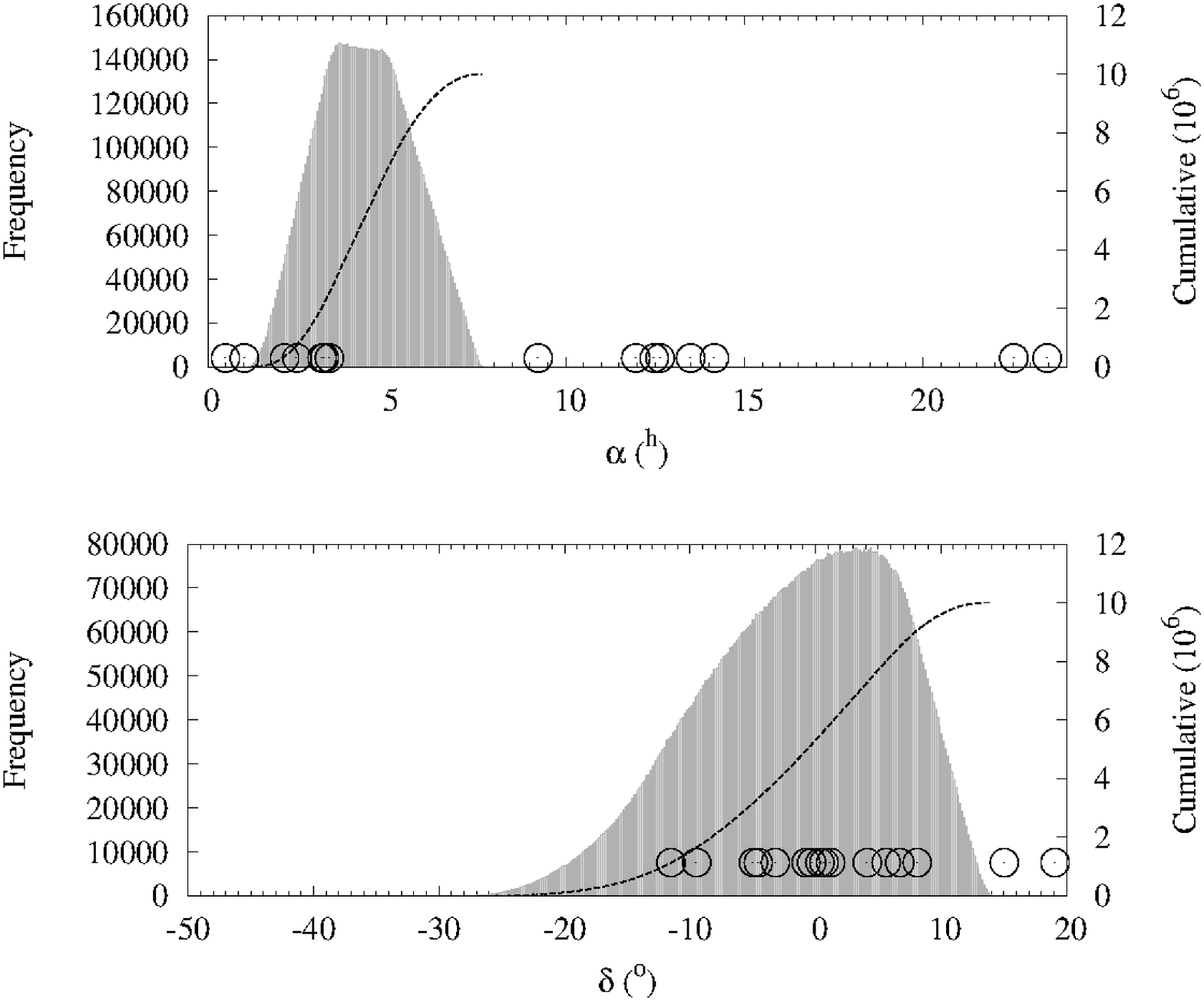}
         \includegraphics[width=0.247\linewidth]{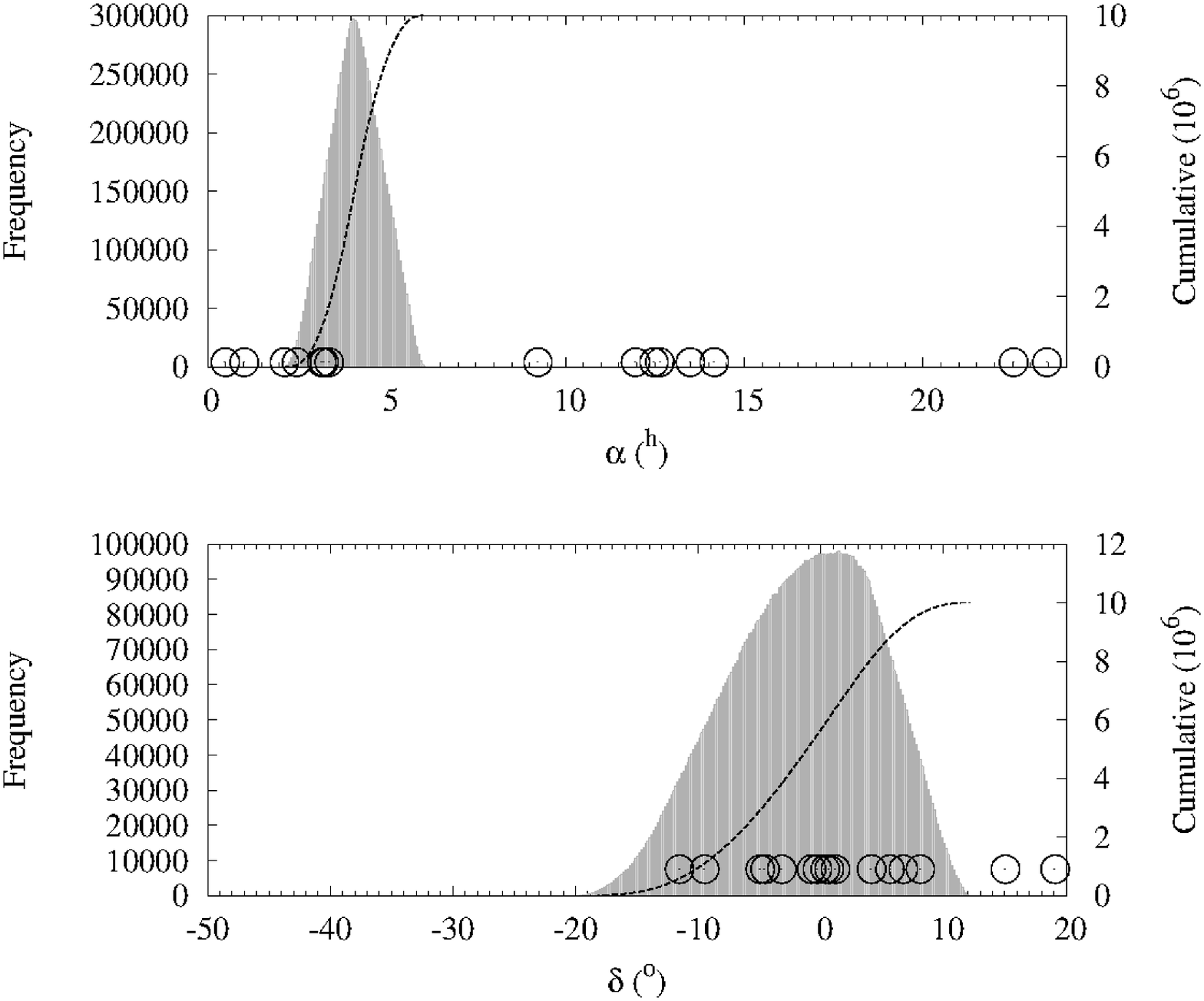}\\
         \caption{Frequency distribution in equatorial coordinates (right ascension, top panel, and declination, bottom panel) of the
                  virtual orbits in Fig. \ref{hunt}. The number of bins is 2 $n^{1/3}$, where $n$ is the number of virtual orbits plotted, 
                  error bars are too small to be seen. The black circles correspond to data in table 2 of de la Fuente Marcos \& de la 
                  Fuente Marcos (2016a).
                 }
         \label{radeca}
      \end{figure*}

        The Planet Nine hypothesis has been further developed in Brown \& Batygin (2016). In this new work, the volume of orbital parameter 
        space linked to the putative planet for an assumed mass of 10~$M_{\oplus}$ is enclosed by $a\in$ (500, 800) au, $e\in$ (0.32, 0.74), 
        $i\in$ (22, 40)\degr, $\Omega\in$ (72.2, 121.2)\degr and $\omega\in$ (120, 160)\degr. Figs \ref{hunt} and \ref{radeca}, second to 
        left-hand panels, show similar trends to the previous ones but now the distribution in equatorial coordinates of the aphelia is 
        wider. The probability is highest in the region limited by $\alpha\in(3, 5)^{\rm h}$ and $\delta\in(-1, 10)\degr$. This enlarges the 
        optimal search area significantly. However, they use a value of $l_q$ of 241\degr$\pm$15\degr that is based on a value of 61\degr 
        for the seven ETNOs so $\Delta{l_q}$ ---not $\Delta\varpi$--- librates about 180\degr; the average value of $l_q$ for these ETNOs 
        from Table \ref{peri} is 62\degr$\pm$38\degr, the average value of $\varpi$ is 60\degr$\pm$39\degr.
%
%
%
%
%
%
%
%

     \subsection{Enforcing apsidal anti-alignment}
        Here, we enforce apsidal anti-alignment and nodal alignment, using the data in Table \ref{bary}, bottom section, and considering the 
        Planet Nine orbital parameter domain defined by $a\in$ (600, 800) au, $e\in$ (0.5, 0.7), $i\in$ (22, 40)\degr, $\Omega\in$ (74.4, 
        137.3)\degr and $\omega\in$ (113.5, 154.4)\degr. Figs \ref{hunt} and \ref{radeca}, second to right-hand panels, show that in this 
        scenario, Planet Nine is most likely to be moving within $\alpha\in(3.0, 5.5)^{\rm h}$ and $\delta\in(-1, 6)\degr$, if it is near 
        aphelion. If the data in Table \ref{bary2} are used instead, then $\Omega\in$ (87, 117)\degr and $\omega\in$ (118.5, 148.8)\degr and 
        we obtain Figs \ref{hunt} and \ref{radeca}, right-hand panels. In this case, the putative planet is most likely moving within 
        $\alpha\in(3.5, 4.5)^{\rm h}$ and $\delta\in(-1, 2)\degr$, i.e. projected towards the separation between the constellations of 
        Taurus and Eridanus. In terms of probability, now the most likely location of Planet Nine is at $\alpha\sim{4}^{\rm h}$ and 
        $\delta\sim0\fdg5$, in Taurus. This area is compatible with Orbit A in de la Fuente Marcos et al. (2016).

  \section{Conclusions}
     In this paper, we have re-analysed the various clusterings in ETNO orbital parameter space claimed in the literature and explored the 
     visibility of Planet Nine within the context of improved constraints. Our results confirm the findings in Batygin \& Brown (2016) and 
     Brown \& Batygin (2016) regarding clustering but using barycentric orbits. However, the observed overall level of clustering may not be 
     maintained by a putative Planet Nine alone, other perturbers should exist. Summarizing: 
     \begin{itemize}
        \item We confirm the existence of apsidal alignment and similar projected pole orientations among the currently known ETNOs. These
              patterns are consistent with the presence of perturbers beyond Pluto and/or, less likely, break-up of large asteroids at 
              perihelion. 
        \item If Planet Nine is at aphelion, it is most likely moving within $\alpha\in(3.0, 5.5)^{\rm h}$ and $\delta\in(-1, 6)\degr$ if  
              $\Delta\varpi\sim$180\degr and $\Delta\Omega\sim$0\degr. 
     \end{itemize}

  \section*{Acknowledgements}
     We thank two anonymous referees for their constructive reports, and S. J. Aarseth, D. P. Whitmire, G. Carraro, D. Fabrycky, A. V. 
     Tutukov, S. Mashchenko, S. Deen and J. Higley for comments on ETNOs and trans-Plutonian planets. This work was partially supported by 
     the Spanish `Comunidad de Madrid' under grant CAM S2009/ESP-1496. In preparation of this paper, we made use of the NASA Astrophysics 
     Data System, the ASTRO-PH e-print server and the MPC data server.

  \newpage
  \appendix
  \section{Further details on the comparison with the results of Brown \& Batygin (2016)}
     According to the heliocentric orbital solutions publicly available from the SBDB, the six (not seven) objects singled out in Batygin \& 
     Brown (2016) all have values of the heliocentric semimajor axis, $a$, larger than 257~au; Brown \& Batygin (2016) indicate that their 
     semimajor axes are larger than 227~au. The six original objects in Batygin \& Brown (2016) ---namely, (90377) Sedna ($a=499$~au), 
     2004~VN$_{112}$ ($a=318$~au), 2007~TG$_{422}$ ($a=483$~au), 2010~GB$_{174}$ ($a=370$~au), 2012~VP$_{113}$ ($a=258$~au) and 
     2013~RF$_{98}$ ($a=307$~au)--- do not have the largest values of the perihelion distance, only four of them ---Sedna ($q=76$~au), 
     2004~VN$_{112}$ ($q=47$~au), 2010~GB$_{174}$ ($q=49$~au) and 2012~VP$_{113}$ ($q=80$~au)--- are in this situation. In Brown \& Batygin 
     (2016), it is said that the seven objects with $a>227$~au are singled out. However, fig. 1(a) in Brown \& Batygin (2016) highlights in 
     red only six objects (those singled out in Batygin \& Brown 2016), not seven; (148209) 2000 CR$_{105}$ ($a=226$~au, $q=44$~au), which 
     has the eighth largest value of the heliocentric semimajor axis and the fifth largest perihelion distance, appears in green in that 
     figure. The ETNO with the seventh largest value of the heliocentric semimajor axis is (82158) 2001 FP$_{185}$ ($a=227$~au, $q=34$~au), 
     not 148209 as erroneously stated in Brown \& Batygin (2016). Our analysis in Figs \ref{cluster} and \ref{poles} shows that 82158 is 
     unlikely to be dynamically connected with the original six ETNOs singled out in Batygin \& Brown (2016), but 148209 very probably is. 
     Therefore, the seven objects of interest within the context of the Planet Nine hypothesis are: Sedna, 148209, 2004~VN$_{112}$, 
     2007~TG$_{422}$, 2010~GB$_{174}$, 2012~VP$_{113}$ and 2013~RF$_{98}$.

     In terms of barycentric (not heliocentric) orbits, see Table \ref{bary}, 148209 has the seventh largest value of the semimajor axis and
     82158 has the eighth largest.

     Table \ref{bary2} shows the statistics for the six original objects in Batygin \& Brown (2016); the average values of the orbital 
     parameters are $e=0.85\pm0.08$, $i=22\degr\pm6\degr$, $\Omega=102\degr\pm33\degr$ and $\omega=314\degr\pm22\degr$ (these values have 
     been used in the discussion in Section 3.1). The average angular separation at pericentre for this group is 45\degr$\pm$34\degr and the 
     average polar separation is 16\degr$\pm$8\degr. For the sample of ETNOs in Batygin \& Brown (2016), both 2007~TG$_{422}$ and 
     2012 VP$_{113}$ are statistical outliers in terms of $e$ (see Tables \ref{bary} and \ref{bary2}).
%
%
      \begin{table*}
        \centering
        \fontsize{8}{11pt}\selectfont
        \tabcolsep 0.10truecm
        \caption{As Table \ref{bary}, bottom section, but only for the six objects singled out in Batygin \& Brown (2016): (90377) Sedna, 
                 2004~VN$_{112}$, 2007~TG$_{422}$, 2010~GB$_{174}$, 2012~VP$_{113}$ and 2013~RF$_{98}$. Source data in Table \ref{bary}, 
                 top section. 
                }
        \begin{tabular}{lrrrrrrrrrrr}
          \hline
             Parameter          & $a$ (au)  & $e$     & $i$ (\degr) & $\Omega$ (\degr) & $\omega$ (\degr) & $\varpi$ (\degr) & $q$ (au) & 
                       $Q$ (au) & $P$ (yr)    & $\Omega^*$ (\degr) & $\omega^*$ (\degr) \\
          \hline
             Mean               & 377.82073 & 0.84595 & 21.88102    & 102.06470        & 313.52297        &  55.58767        & 54.05935 &
                      701.58211 &  7504.04034 &    102.06470      &  $-$46.47703      \\ 
             Std. dev.          & 101.95284 & 0.07978 &  6.12694    &  32.70823        &  22.38639        &  40.68277        & 19.59205 &
                      206.62725 &  3042.77393 &     32.70823      &     22.38639      \\
             Median             & 339.28128 & 0.85859 & 22.80700    & 101.85731        & 313.83049        &  35.80781        & 47.94245 &
                      630.62012 &  6248.20658 &    101.85731      &  $-$46.16951      \\
             Q$_{1}$            & 319.65774 & 0.85096 & 19.33709    &  73.35133        & 297.98366        &  26.52012        & 39.04232 &
                      600.27317 &  5711.85155 &     73.35133      &  $-$62.01634      \\
             Q$_{3}$            & 464.31370 & 0.87960 & 25.17359    & 126.26351        & 324.34156        &  81.41511        & 69.28395 &
                      865.41241 & 10075.01110 &    126.26351      &  $-$35.65844      \\
             IQR                & 144.65595 & 0.02864 &  5.83650    &  52.91217        &  26.35790        &  54.89498        & 30.24163 &
                      265.13924 &  4363.15955 &     52.91217      &     26.35790      \\
             OL                 & 102.67381 & 0.80799 & 10.58233    & $-$6.01693       & 258.44681        & $-$55.82235      &$-$6.32013 &
                      202.56430 & $-$832.88777 & $-$6.01693    & $-$101.55319         \\
             OU                 & 681.29763 & 0.92256 & 33.92835 & 205.63177           & 363.87841        & 163.75758        &114.64641 &
                     1263.12127 & 16619.75042 &  205.63177        &    3.87841        \\
          \hline
        \end{tabular}
        \label{bary2}
      \end{table*}
%
%

     The expression for the eccentricity of the putative Planet Nine in Brown \& Batygin (2016) produces negative values for masses under
     $\sim$6.38~$M_{\oplus}$ and the lower limit of the value of the semimajor axis. Production of unphysical values is the main reason why
     it has not been applied to select the range in eccentricities used in Section 3.3.

  \bsp
  \label{lastpage}
\end{document}